\documentclass[a4paper]{jpconf}
\usepackage{graphicx}
\newcommand{\be}{\begin{equation}}
\newcommand{\ee}{\end{equation}}
\newcommand{\ba}{\begin{eqnarray}}
\newcommand{\ea}{\end{eqnarray}}
\newcommand{\ban}{\begin{eqnarray*}}
\newcommand{\ean}{\end{eqnarray*}}

\newcommand{\bef}{\begin{figure}}
\newcommand{\eef}{\end{figure}}
\newcommand{\bce}{\begin{center}}
\newcommand{\ece}{\end{center}}
\begin{document}
\title{Exciting  gauge unstable modes  of the quark-gluon plasma by relativistic jets}

\author{Massimo Mannarelli and Cristina Manuel}

\address{Instituto de Ciencias del Espacio (IEEC/CSIC),
Campus Universitat Aut\`onoma de Barcelona, Facultat de Ci\`encies, Torre C5 E-08193 Bellaterra (Barcelona), Spain}

\ead{massimo@ieec.uab.es}

\begin{abstract}
We present  a study of the
 properties of the collective modes of a system composed by a thermalized quark-gluon plasma traversed by a relativistic jet of partons. We find that when the  jet traverses the system unstable gauge field modes are excited and  grow on very short time scales. 
The aim is to provide a novel mechanism for the description of the jet quenching  phenomenon, where the jet crossing the plasma loses energy exciting colored unstable modes. In order to simplify the  analysis   we employ a  linear response approximation,  valid for short time scales.  We assume  that the  partons in the jet  can be described with a tsunami-like distribution function, whereas we treat the quark-gluon plasma employing two different approaches. In the first approach we adopt a Vlasov approximation for the kinetic equations, in the second approach we solve a set of fluid equations.  In both cases we derive the expressions of the dispersion law of the collective unstable modes and compare the results obtained.   
\end{abstract}

\section{Introduction}
One of the methods for unveiling the properties of matter produced in ultrarelativistic heavy-ion
collisions is to study the propagation  properties of high $p_T$ partons generated  by hard scatterings in the initial stage of the collision. 
When the jet of partons travels across the medium it loses  energy  and degrades, mainly by radiative processes (see \cite{Kovner:2003zj} for reviews). The energy and momentum of the jet are   absorbed   by the plasma and result in an increased  production of  soft hadrons in the direction of propagation of the partons.

We  propose a novel mechanism \cite{Mannarelli:2007gi,Mannarelli:2007hj} for describing how the jet loses energy and momentum while traveling in a thermally equilibrated   quark-gluon plasma (QGP).  
Since the jet of particles  is  not in thermal equilibrium with the QGP it perturbs and destabilizes the system  inducing the generation of gauge fields.  Some of these gauge modes are unstable and grow  exponentially fast in time absorbing the kinetic energy of the  jet.

The study of the interaction of a relativistic stream of particles with a  plasma is a topic of interest in different fields of physics, ranging from inertial confinement fusion,
astrophysics and cosmology. When the particles of the stream are charged, plasma instabilities develop,
leading to an initial stage of fast growth of the gauge fields. 
The study of chromo instabilities is  a very active field of research 
(see \cite{Mrowczynski:2006ad} for a review).

In order to study the system composed by the jet and the plasma  we have employed two different methods. 
In \cite{Mannarelli:2007gi}    both the plasma and the jet  are described using a fluid approach. This  approach developed in \cite{Manuel:2006hg} has been derived from kinetic theory expanding the transport equations in moments of momenta and truncating the expansion at the second moment level. The system of equations is then closed with an equation of state relating  pressure and energy density. 
The fluid  approach  has several advantages with respect to the underlying kinetic theory. The most remarkable one is that  one has to deal with  a set of  equations much simpler than those of kinetic theory. Then one can    easily generalize the fluid equations  to  deal with more complicated systems. This is a strategy that has been successfully followed in the  study of
different dynamical aspects of non-relativistic electromagnetic plasmas \cite{Kra73}.

In the second approach  we consider  the same setting,  {\it i.e.}  a static plasma traversed by a relativistic jet, but  we use kinetic theory  instead of the fluid approach. Transport theory provides a well controlled framework for studying the properties of the quark-gluon plasma in the weak coupling regime, $g \ll 1$.
Indeed it is well known that the physics of long distance scales in an equilibrated weakly coupled
QGP can be described within semiclassical transport equations \cite{Braaten:1989mz,Bla93,Kel94}. In this approach the
hard modes,  with typical energy scales of order $T$,  are treated as (quasi-)particles which propagate in the background of the soft modes, whose energies are equal or less than $gT$, which are treated as classical gauge fields.
This program has been very successful for  understanding some dynamical aspects of the soft gauge fields in an almost equilibrated QGP \cite{Blaizot:2001nr,Litim:2001db}.

We conclude  comparing the results for the growing rates of the unstable modes  obtained with kinetic theory in Ref.~\cite{Mannarelli:2007hj} with the analogous results we obtained with the fluid approach in Ref.~\cite{Mannarelli:2007gi}.

\section{Kinetic theory approach}
We consider a system composed by a quark-gluon plasma traversed by a jet of partons. We  assume that  the system is initially in a  colorless and thermally equilibrated  state and we will study the behavior of  small deviation  from  equilibrium. The distribution functions of  quarks, antiquarks and gluons   belonging to the quark-gluon plasma are, respectively
\begin{equation}
Q(p,x) = f^{\rm eq.}_{FD}(p_0) + \delta Q(p,x) \,,\,\,\,\,  \bar Q(p,x) = f^{\rm eq.}_{FD}(p_0) + \delta \bar Q(p,x) \,,\,\,\,\, 
G(p,x) = f^{\rm eq.}_{BE}(p_0) + \delta G(p,x) 
\end{equation}
where  the various quantities are hermitian matrices
in color space (we have suppressed color indices) and where 
$f^{\rm eq.}_{FD/BE}(p_0) = \frac{1}{e^{p_0/T} \pm 1}$
are the (colorless) Fermi-Dirac and Bose-Einstein equilibrium distribution functions.

Also the particles  of the jet are assumed to be initially in equilibrium and that
small color fluctuations are present: 
\begin{equation}
W_{\rm jet}(p,x) = f_{\rm jet} (p) + \delta W_{\rm jet}(p,x) \,.
\end{equation}
For the initial jet distribution function we will not consider a thermal distribution function. We will indeed approximate the distribution function of the jet with a colorless tsunami-like form \cite{Pisarski:1997cp}
\begin{equation}
\label{tsunami}
f_{\rm jet}(p) =  \bar n \, \bar u^0 \;
\delta^{(3)}\Big({\bf p}
- \Lambda \, \bar {\bf u} \Big) \;,
\end{equation}
that describes a system of particles of   constant density $\bar n$, all moving with the same  velocity $\bar u^\mu = (\bar u^0, \bar {\bf u}) = \gamma (1, {\bf v})$, where $\gamma$ is the Lorentz factor and   $\Lambda$ fixes the scale of the energy of the particles. 

Although this distribution function is adequate for describing a uniform and sufficiently dilute system of particles, it would be very interesting to extend our analysis to more complicated  forms. Using more involved distribution functions   on the one hand  would lead to a more accurate description of the jets  relevant for heavy ion phenomenology, 
when the density of particles composing the jet is not uniform and there is a spread in momentum. However, on the other hand this would  complicate the study  of the collective modes. 

The distribution function of quarks satisfy the following transport equation 
\begin{eqnarray}
p^{\mu} D_{\mu}Q(p,x) + {g \over 2}\: p^{\mu}
\left\{ F_{\mu \nu}(x), \partial^\nu_p Q(p,x) \right\}
&=& C \;,
\label{transport-eq}
\end{eqnarray}
whereas  the distribution functions of antiquarks, gluons and of the particles of the jet satisfy similar equations. With  $\{...,...\}$ we denote the anticommutator, $\partial^\nu_p$ is
the four-momentum derivative and $g$ is the QCD coupling constant.  The covariant derivatives $D_{\mu}$ and ${\cal D}_{\mu}$ act as
$$
D_{\mu} = \partial_{\mu} - ig[A_{\mu}(x),...\; ]\;,\;\;\;\;\;\;\;
{\cal D}_{\mu} = \partial_{\mu} - ig[{\cal A}_{\mu}(x),...\;]\;,
$$
with $A_{\mu }=  A^{\mu }_a (x) \tau^a$ and ${\cal A}_{\mu }=  A^{\mu }_a (x) T^a$, and $\tau^a$ and $T^a$ are $SU(3)$ generators in the fundamental and adjoint representations, respectively.
The strength tensor in the fundamental representation is
$F_{\mu\nu}=\partial_{\mu}A_{\nu} - \partial_{\nu}A_{\mu}
-ig [A_{\mu},A_{\nu}]$, while  ${\cal F}_{\mu \nu}$ denotes the field
strength tensor in the adjoint representation. 

In Eq.(\ref{transport-eq})  $C$ represents the collision term. For time scales shorter than the mean free path time the collision terms  can be neglected, as typically done in the so-called Vlasov approximation. 
The knowledge of the distribution function allows one to compute the associated color current, which in a self-consistent
treatment enters as a source term in the Yang-Mills equation. 
Therefore in this approximation  the different components of the system formed by the plasma and the jet interact with each other only through the generated average gauge fields.

In the Vlasov approximation one can compute the contribution to the  polarization tensor
of  the particles of species $\alpha$ (where $\alpha$ refers to quarks, antiquarks, gluons or
the partons of the jet)
 as
\begin{equation}
\label{Pi-kinetic}
\Pi^{\mu \nu}_{ab, \alpha}(k) = - g^2 C^{\alpha}_F \delta_{ab}\int_p
f_{\alpha} (p) \;
{ (p\cdot k)(k^\mu p^\nu + k^\nu p^\mu) - k^2 p^{\mu} p^{\nu}
- (p\cdot k)^2 g^{\mu\nu} \over(p\cdot k)^2} \;,
\end{equation}
where $a,b$ are color indices and $C^\alpha_F$ is the value of the quadratic Casimir associated with the particle specie $\alpha$ which takes values
$1/2$ and $3$ for the fundamental and adjoint representations, respectively.
The momenta measure is defined as
\begin{equation}
\label{measure}
\int_p \cdots \equiv \int \frac{d^4 p}{(2\pi )^3} \:
2 \Theta(p_0) \delta (p^2-m^2_\alpha) \;,
\end{equation}
where $m_\alpha$ is the mass of the particle of specie $\alpha$.
For simplicity we  assume that the particles belonging to the plasma are massless.  Instead the particles of  the jet  have a non-vanishing mass.

When $f_\alpha(p)$ is a thermal equilibrated distribution function, Eq.~(\ref{Pi-kinetic})
reduces to the form of the hard thermal loop (HTL) polarization tensor \cite{Braaten:1989mz,Bla93,Kel94}.  However  for the tsunami-like distribution function, Eq.~(\ref{tsunami}),
the polarization tensor obviously takes a different form.

The gauge fields  obey the Yang-Mills equation 
\begin{equation}
\label{yang-mills1}
D_{\mu} F^{\mu \nu}(x) = \delta j^\nu_t (x) = \delta j_{p}^{\nu}(x) + \delta j_{\rm jet}^{\nu }(x)\; ,
\end{equation}
where we have defined
\begin{equation}
\label{col-current}
\delta j^{\mu }_p(x) = -\frac{g}{2} \int_p p^\mu \;
\Big[ \delta Q( p,x) - \delta \bar Q ( p,x)+  2 \tau^a {\rm Tr}\big[T^a \delta G(p,x) \big]\Big] \;,
\end{equation}
which describes the plasma color current, and
\begin{equation}
\label{jetcol-current}
\delta j_{\rm jet}^{\mu }(x) = -\frac{g}{2} \int_p p^\mu \;
 \delta W_{\rm jet}( p,x)  ,
\end{equation}
which describes the fluctuations of the  current associated with the jet.

Equation (\ref{yang-mills1}) together with Eq.~(\ref{transport-eq})
form a set of equations that has to be solved  self-consistently. Indeed the gauge fields which are solutions of the Yang-Mills
equation enter into the  transport equations of every particle species and,  in turn, affect the evolution of the distribution functions.

\section{Liquid approach}
Hydrodynamical equations are the
expressions of the conservation laws of a system when it is in
local equilibrium. In Ref.~\cite{Manuel:2003zr} the local
equilibrium state for the quark-gluon plasma has been determined.
It is in general described by one singlet four velocity, a baryon
density, singlet energy and pressure, and in principle, a
non-vanishing color density. However, dynamical processes
associated to the existence of Ohmic currents tend to whiten the
plasma quickly, on time scales much shorter than  momentum
equilibration processes \cite{Manuel:2004gk}. For this  reason one
can expect that only  colorless (singlet) fluctuations are relevant at large time and space scales.
However, there are situations,  as the one considered here, when color fluctuations grow on short time scales instead of being damped. Therefore in order to describe  the short time evolution of the plasma, one needs to include color hydrodynamical fluctuations in the equations.  
In  Ref.~\cite{Manuel:2006hg} such a  chromohydrodynamical approach  for  the short time evolution
of the system has been formulated. The fluid equations can be obtained expanding 
the collisionless transport Equation~(\ref{transport-eq}) (and the analogous equations for antiquarks and gluons) in moments of momenta and truncating the expansion at the second order level. For simplicity, as in Ref.~\cite{Manuel:2006hg}, we will only consider
the contribution of quarks in the fundamental representation. The inclusion of antiquarks and gluons is
straightforward. The fluid approach consist of the covariant continuity equation for the fluid four-flow
\begin{equation}\label{cont-eq}
D_\mu n^\mu  =  0 \;
 \end{equation}
and of the equation that couples the energy-momentum tensor
$T^{\mu \nu}$ to the gauge fields 
\begin{equation}\label{en-mom-eq} 
D_\mu
T^{\mu \nu} - {g \over 2}\{F_{\mu}^{\;\; \nu}, n^\mu \}= 0 \;. \end{equation}

We further assume that the four-flow and
the energy-momentum tensor have the expression valid for an ideal fluid, i.e.
\begin{equation}
\label{flow-id}
n^\mu(x)  = n (x) \, u^\mu(x) \hspace{.8cm} {\rm and} \hspace{.5cm}
T^{\mu \nu}(x)  = {1 \over 2}
\big(\epsilon (x) + p (x)\big)
\big\{u^\mu (x), u^\nu (x) \big\}
- p (x) \, g^{\mu \nu}\;,
\end{equation}
where the hydrodynamic velocity $u^\mu$,  the particle density
$n$, the energy density $\epsilon$ and the pressure $p$ are $3 \times 3$
matrices in color space.  

The color current due to the flow of the fluid  can be expressed in terms of the hydrodynamic velocity  and  the particle density as
\begin{equation}
\label{hydro-current}
j^\mu(x) = -\frac{g}{2}
\Big(n u^\mu - {1 \over 3}{\rm Tr}\big[n
u^\mu  \big]\Big) \;. \end{equation}
As in the kinetic theory approach, the color current acts as a source term
for the gauge fields in the Yang Mills equation
\begin{equation}
\label{yang-mills2}
D_{\mu} F^{\mu \nu}(x) = j^{\nu}(x)\; .
\end{equation}
Thus,  we will assume that all the gauge fields that appear in the fluid equations
are  only due to the presence of a colored current in the medium.
The fluctuations of the current induced by the fluctuation of the density and of the hydrodynamic velocity  are related in linear response theory to fluctuations of the gauge fields via \be
\delta j^\mu_a(k) = -
\Pi^{\mu \nu}_{ab} (k) A_{\nu,b}(k) \;, \ee
which defines the polarization tensor in the fluid approach. 

Notice that  Eqs.~(\ref{cont-eq}), (\ref{en-mom-eq}) and  (\ref{yang-mills2}) do not form a closed set  and one more relation has to be provided. In hydrodynamical treatments, one usually imposes a relation between pressure and energy density.  We will assume that pressure and energy density fluctuations are related by $\delta p^a(x) = (c_s^a)^2 \, \delta \epsilon^a(x)$ where  $c_s^a$ is a parameter.

Also  notice  that since Eqs.~(\ref{cont-eq}) and (\ref{en-mom-eq}) were derived  from the collisionless transport Eq.~(\ref{transport-eq}) 
obeyed by the particle distribution function,  the conservation laws expressed by the Eqs.~(\ref{cont-eq}) and (\ref{en-mom-eq})  are
strictly valid on time scales shorter than the mean free path time.

\section{Collective modes in the system composed by the QGP and jet}
\label{QGP-JET}

We now consider the collective modes of the system composed by  an equilibrated QGP traversed by a jet of particles. We are interested in very short time scales when the Vlasov approximation can be employed. The  effect of the beam of particles is to induce a color current, which provides a contribution to the polarization tensor. The polarization tensor of the whole system is additive in this short time regime, meaning that
\be
\Pi^{\mu \nu}_{t}(k) = \Pi^{\mu \nu}_{p}(k) + \Pi^{\mu \nu}_{\rm jet}(k) \, ,
\ee
where $\Pi^{\mu \nu}_{p}(k)$ and  $\Pi^{\mu \nu}_{\rm jet}(k)$ are the polarization tensor of the plasma and of the jet respectively. The total dielectric tensor is given by
\be
\label{total-dielectric}
 \varepsilon^{ij}_{\rm t}(\omega,{\bf k}) = \delta^{ij} + \frac{\Pi^{ij}_t}{\omega^2}  \,,
\ee
and the dispersion laws of the collective modes of the whole system  can  be determined solving  the equation
\be
\label{dispersion-T}
 {\rm det}\Big[ {\bf k}^2 \delta^{ij} -k^i  k^j
- \omega^2 \varepsilon^{ij}_{\rm t}(k)  \Big]  = 0 \,.
\ee

The solutions of this equation depend on ${|\bf k|}$, ${|\bf v|}$,
$\cos\theta={\bf \hat k \cdot \hat v}$, $\omega_t^2 = \omega^2_{\rm p}+\omega^2_{\rm jet}$ and on $b = \frac{\omega^2_{\rm jet}}{\omega^2_{t}}$. Clearly, when the plasma and the jet do not interact (or equivalently $b=0$), they have stable collective modes. However, once we consider the composed system of plasma and jet interacting via mean gauge field interactions, unstable gauge modes may appear. 

\begin{figure}[h]
\begin{center}
\includegraphics[width=14pc]{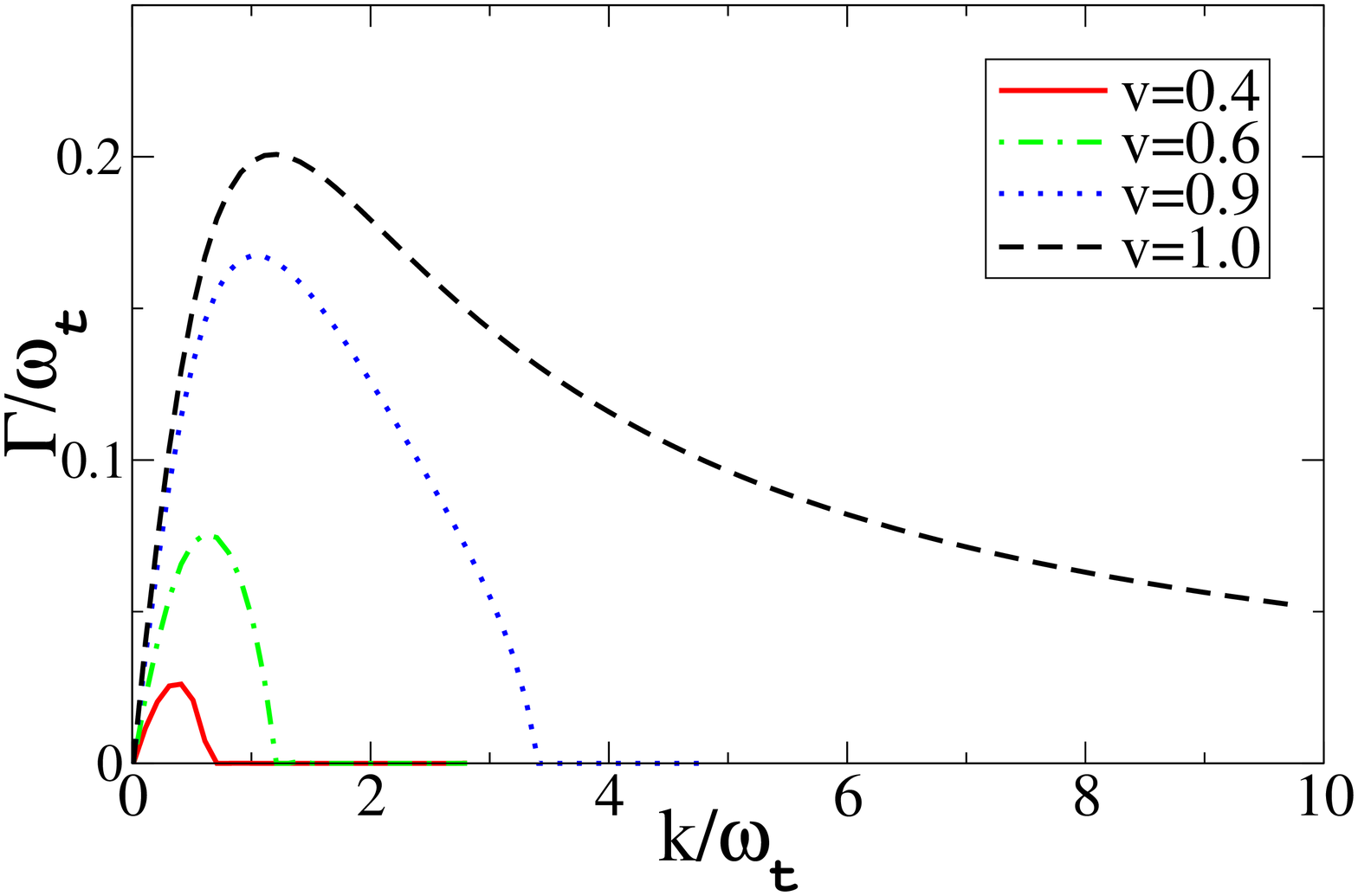}
\includegraphics[width=14pc]{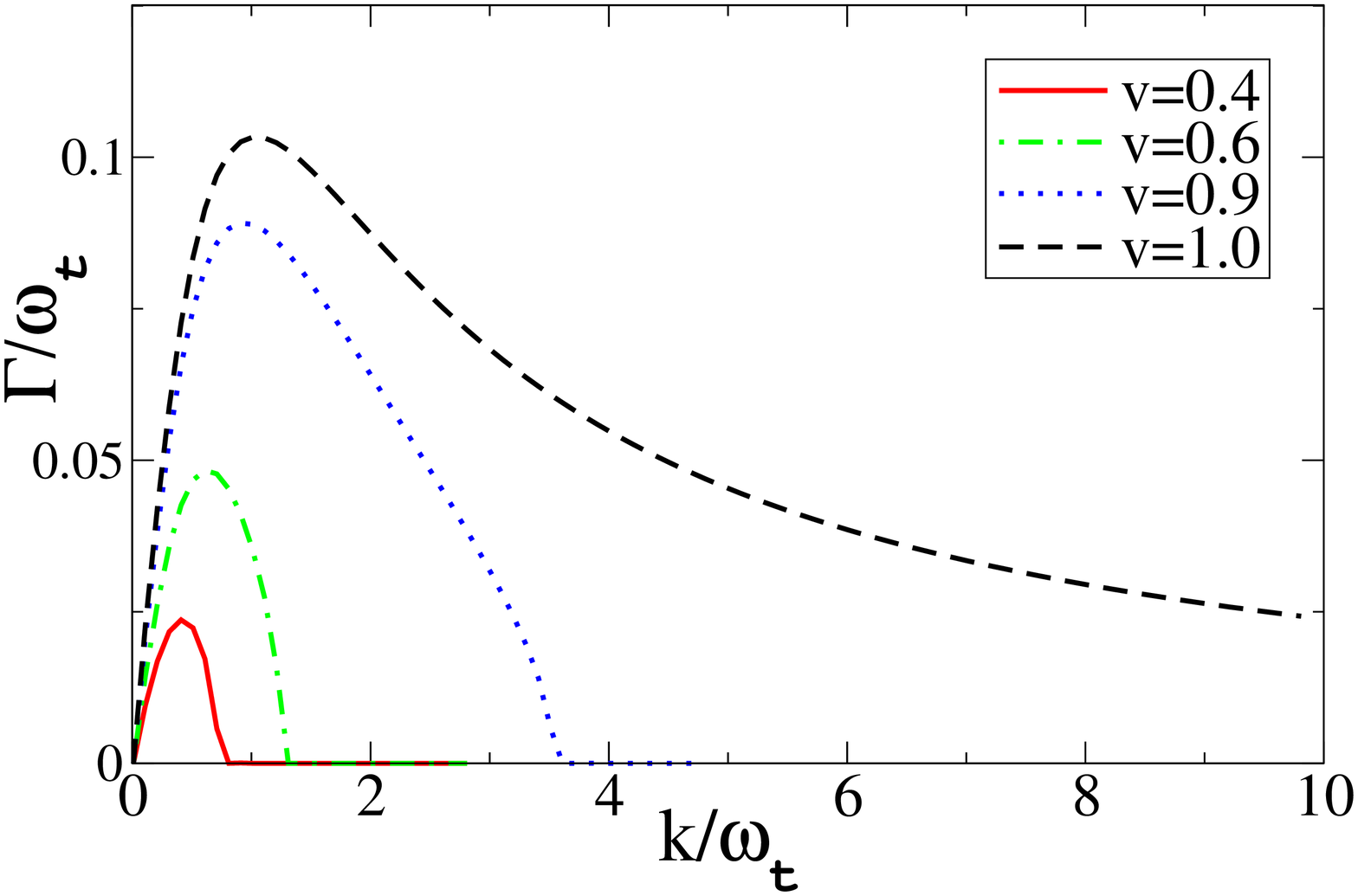}
\caption{\label{perp} Imaginary part of the dispersion law of the unstable  mode  for the system composed by a plasma and a jet with the kinetic theory approach for  $\bf k \perp v$ as a function of the momentum of the mode.  Left panel refers to $b=0.1$   and right panel refers to $b=0.02$. In both case results for four different values of the velocity of the jet,  $|{\bf v}|$, are shown.}  \label{Orthofig1}
\end{center}
\end{figure}

In Fig. \ref{perp} we report the results for the unstable mode obtained with the kinetic theory approach for the case where ${\bf k} \perp  {\bf v} $. The case ${\bf k} \perp  {\bf v} $ corresponds to the most unstable mode in the kinetic theory approach. Estimating the maximum growing rate in the weak coupling limit  at $T \sim 350$ MeV, we find that instabilities develop on time scales $t \sim 1-2$ fm/c.

\begin{figure}[h]
\begin{center}
\includegraphics[width=14pc]{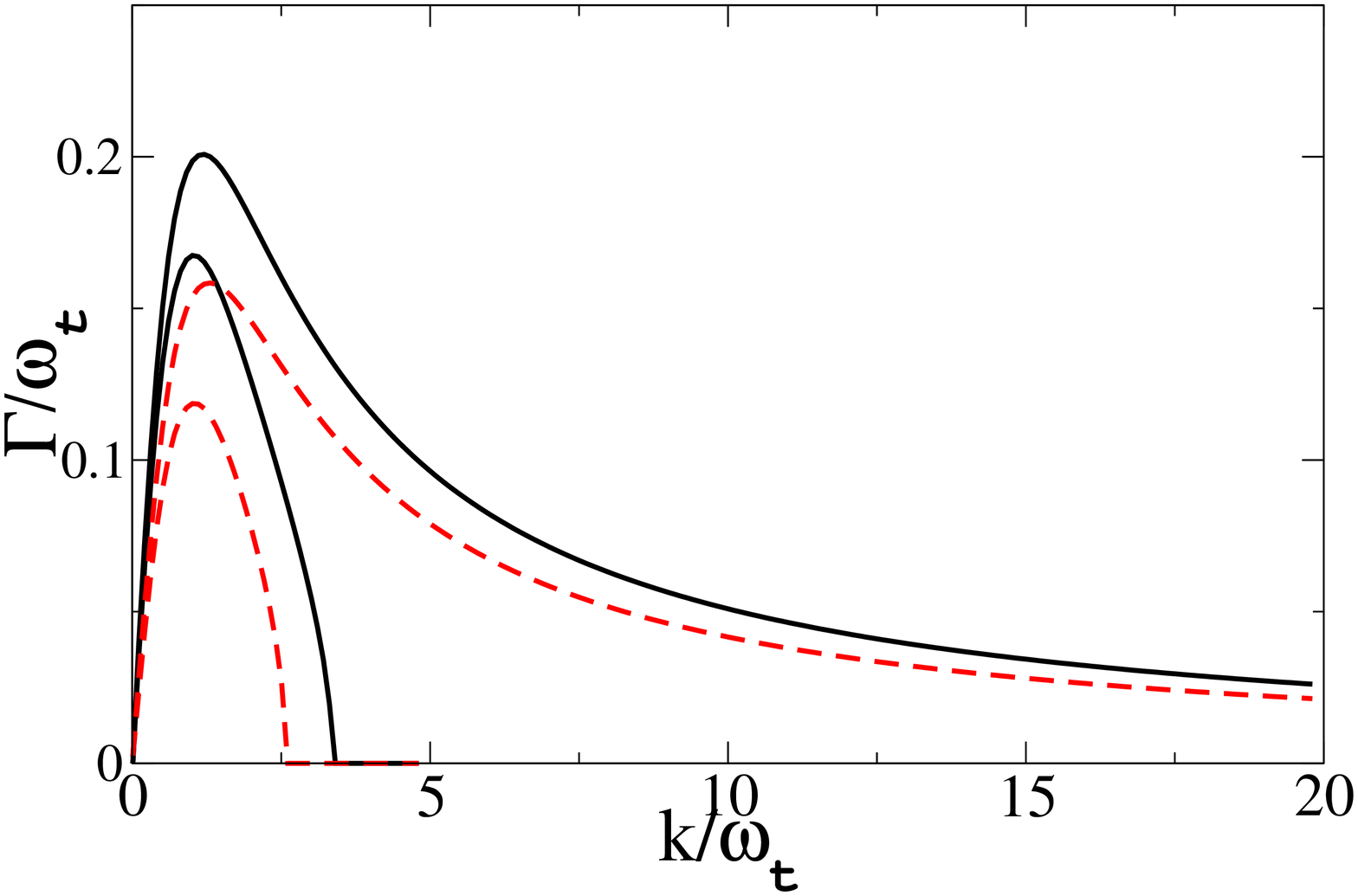}
\includegraphics[width=14pc]{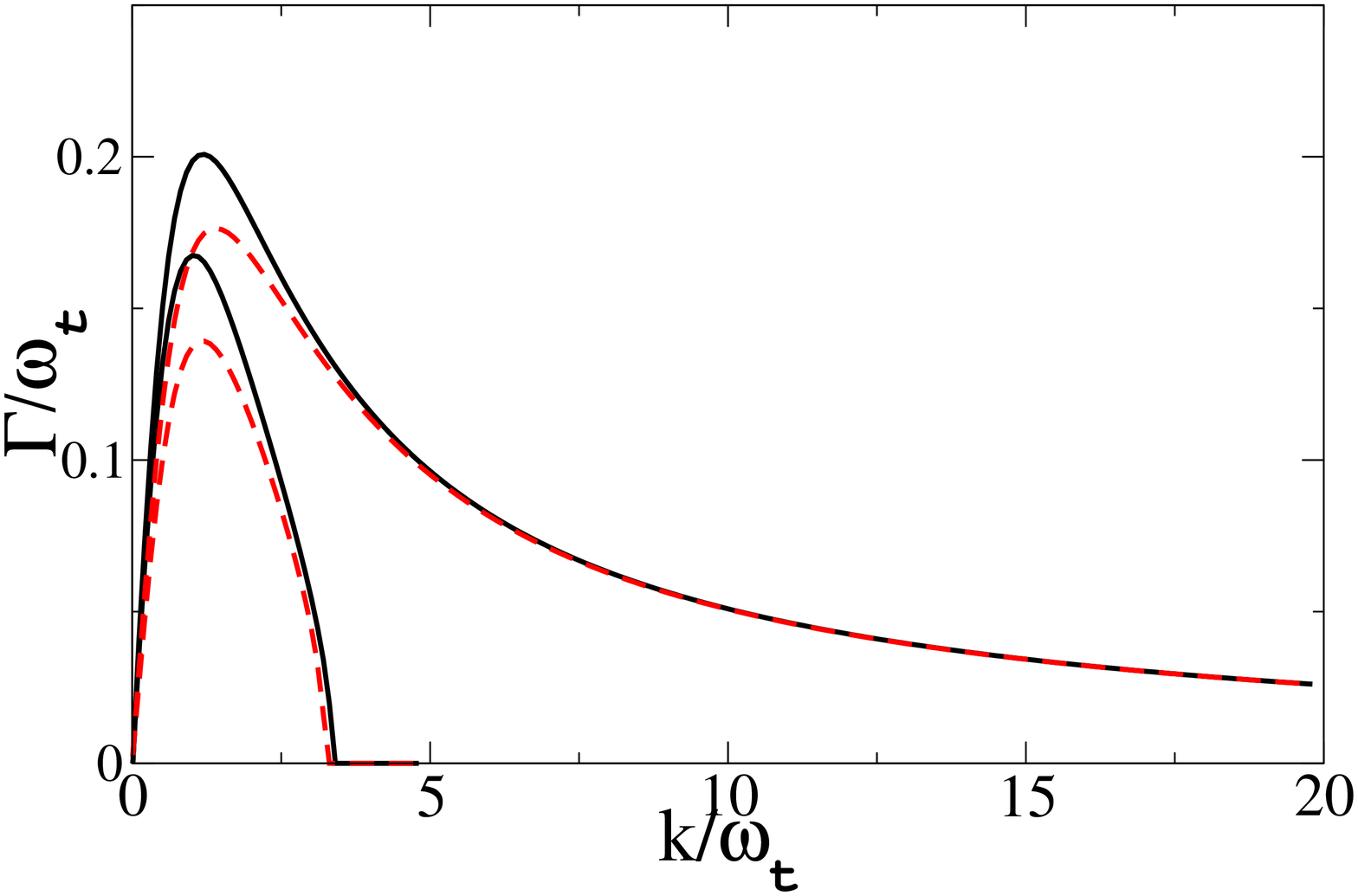}
\caption{\label{comortofig} Comparison between the imaginary part of the dispersion law of the unstable  mode in the two approaches in the case $\bf k \perp v$ as a function of the momentum of the mode at  $b=0.1$ for $v=0.9$ (lower curves) and $v=1.0$ (upper curves) in the conformal limit, $c_s^a= 1/\sqrt{3}$,  (left panel) and for $c_s^a= 1/2$ (right panel). Dashed (red) lines correspond to the results obtained with the fluid approach;  full (black) lines  correspond to results obtained with kinetic theory.} 
\end{center}
\end{figure}

In Fig. \ref{comortofig} we compare the results for the transverse unstable modes obtained with the kinetic theory with the analogous results obtained within the fluid approach. The agreement between the two approaches is quite remarkable. However, it depends on the value of the parameter $c_s^a$. On the left panel we have chosen the ``conformal" value  $c_s^a=1/\sqrt{3}$. On the right panel we have employed  $c_s^a =1/2 $, that is the value that minimizes the difference between the results of the two approaches  in this case \cite{Mannarelli:2007hj}.  

There are two qualitative differences between the results obtained within the kinetic theory method and the fluid  approach \cite{Mannarelli:2007hj}. First, in the fluid approach the instabilities develop for velocities $v>c_s^a$ whereas in  kinetic theory the threshold value of the velocity for the development of the instability is not related to the parameter
$c_s^a$. This is  due to the fact that the equation of state  does not enter the kinetic theory picture and therefore  $c_s^a$ does not play any role. 

Second, in the Vlasov approximation for sufficiently small velocities, $v < 0.6 $, there is no preferred unstable direction, whereas for larger velocities   the most unstable  modes correspond to large angles between $\bf k$ and $\bf v$. This has to be contrasted with  the results obtained using fluid equations, where one finds that for velocities $v \sim  c_s^a$  the most unstable mode correspond to momenta $\bf k$ collinear with the velocity of the jet, whereas for  ultrarelativistic velocities  $v \sim 1  $ the unstable modes corresponding to angles $\theta > \pi/8 $  are dominant and the most unstable mode corresponds to  $\theta \simeq \pi/4 $.

\section*{Acknowledgments}
This work has been supported by the Ministerio de Educaci\'on y Ciencia (MEC) under grant AYA 2005-08013-C03-02.

\section*{References}


\end{document}